\documentclass[twocolumn,prl,showpacs,floatfix]{revtex4}
\usepackage{graphicx}
\usepackage{bm}
\begin{document}
\title{Shot Noise in Ballistic Quantum Dots with a Mixed Classical Phase Space}
\author{H.-S. Sim}
\author{H. Schomerus}
\affiliation{
Max-Planck-Institut f\"{u}r Physik komplexer Systeme, 
N\"{o}thnitzer Str. 38, 01187 Dresden, Germany
}
\date{\today}
\begin{abstract}
We investigate shot noise for
quantum dots whose classical phase space
consists of both regular and chaotic regions.
The noise is systematically suppressed below the
universal value of fully chaotic systems, by an amount which varies with the
positions of the leads.
We analyze the dynamical origin of this effect by a novel way to
incorporate diffractive impurity scattering.
The dependence of the shot noise
on the scattering rate shows that 
the suppression arises
due to the deterministic nature of transport through
regular regions and along short
chaotic trajectories.
Shot noise
can be used to probe phase-space structures of
quantum dots with generic classical dynamics.
\end{abstract}


\maketitle

The phenomenon of
shot noise, the time-dependent fluctuations in electrical currents
caused by the discreteness of the electron charge $e$,
has been extensively studied in recent years in mesoscopic systems
(for a review see
\cite{Blanter_Phyrep}).
The focus of interest 
was on classically chaotic ballistic quantum dots
(electron billiards) and on diffusive
quantum wires, which can be investigated
theoretically by applying random matrix theory
\cite{wire1a,Jalabert,Beenakker_RMP}
as well as semiclassical methods \cite{wire1b,deJong_physica,Blanter_PRL}.
For incoherent transport through
a chaotic quantum dot the shot noise would assume the
Poissonian value $P_0 =2 e G_0 V$, where $G_0=N e^2/ (2 h)$ is the
serial conductance of the two quantum point contacts which
connect the dot to electronic reservoirs (maintained
at a voltage difference $V$) by $N$ channels.
For low temperatures
correlations of electrons due to Fermi 
statistics suppress the noise $P$ by a factor of ${\cal F}=P/P_0$
relative to this value
of uncorrelated electrons.
In chaotic quantum dots
the suppression factor ${\cal F}={\cal F}_{\rm ch}=1/4$
is universal \cite{Jalabert},
{\it i.\,e.}, independent of the details
of the system, which also has been confirmed by an experiment
\cite{Oberholzer}.
The origin of the low-temperature noise is the {\em probabilistic} nature of
quantum transport, arising from attempts to transmit charge carriers
between electronic reservoirs with a finite success probability $\in [0,1]$.
Very recently, non-universal corrections to the shot-noise due
to residual signatures of  classically
{\em deterministic} scattering \cite{Beenakker_semi}  have been discussed
by Agam, Aleiner, and Larkin \cite{Agam}.
They
predicted that
shot noise in a chaotic dot can be further reduced below ${\cal
F}_{\rm ch}$ under the condition that 
electrons pass the dot without sufficient diffraction. This has been
verified in a recent experiment \cite{Oberholzer2}.

In this paper we address the shot noise of
{\em generic} ballistic quantum dots. They are not classically chaotic
\cite{Markus} but
possess a mixed phase space,
where regular islands are separated from chaotic seas by impenetrable
dynamical barriers. 
Signatures of the mixed phase space in quantum transport
have been found in the conductance,
which exhibits fractal fluctuations \cite{Ketzmerick} 
and isolated resonances \cite{Huckestein}.
Shot noise can be seen as the second cumulant of charge counting
statistics, with the conductance being first cumulant.
As we will demonstrate, shot noise carries valuable dynamical information
which can be extracted systematically.
For generic quantum dots the suppression factor ${\cal F}$ is found to be
systematically reduced
below the universal value ${\cal F}_{\rm ch}$ for fully 
chaotic systems, not only due to the deterministic nature of transport
along short chaotic trajectories (this mechanism of Ref.\ \cite{Agam}
will be confirmed), but also because quantum diffraction
is strongly reduced for transport through regular regions, as well.

Our investigation of the dynamical origin of the additional shot-noise
reduction is based on a novel procedure to analyze
shot noise (which we obtain from a numerical simulation)
with help of the Poisson kernel \cite{Mello,Brouwer,Baranger_Poisson},
a statistical ensemble of random
matrix theory \cite{Beenakker_RMP}.
In the Poisson kernel one averages the scattering matrix over an 
energy range $E_{\rm av}$, in this way eliminating 
the system-specific details of the dynamics 
with
time scales longer than $t_{\rm av}=\hbar / E_{\rm av}$,
and replaces these by random dynamics of the same universality class as
elastic diffractive impurity scattering (equivalently, fully chaotic dynamics).
(Ref.\ \cite{Agam} also employed diffractive impurity scattering; however,
there it was used 
to mimic the non-deterministic processes of quantum chaos
within a theory that cannot account for them, while in our case
all quantum effects are fully included from the beginning.)
The effective mean free scattering
time $t_{\rm av}$ can be tuned by changing
the energy-averaging window $E_{\rm av}$.
The dependence of the suppression factor ${\cal F}$ on
$t_{\rm av}$ then allows to
analyze the properties of trajectories with
classical dwell times $t_{\rm dwell}\simeq t_{\rm av}$
(up to a possible factor of order unity),
over the large range of dwell times
that is typically involved in
the classical transport through generic quantum dots (in contrast,
chaotic dots are characterized  only by
a single time scale, the mean dwell time).
The analysis via the Poisson kernel is supplemented by a
semiclassical estimation of the shot noise,
obtained from classically deterministic motion course-grained over
Planck cells while preserving Fermi statistics, in the spirit of Refs.\
\cite{deJong_physica,Blanter_PRL}.
It follows that shot noise is suppressed most
strongly if the leads are well-coupled
to classical regular regions 
(with area larger than a Planck cell).

\begin{figure}
\includegraphics[width=0.40\textwidth,height=0.22\textheight]{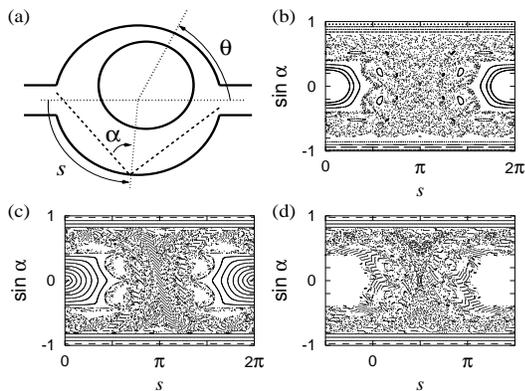}
\caption{
(a) Schematic diagram of the annular billiard between
two circular hard walls of exterior radius $R$, interior radius
$r=0.6 R$, and eccentricity
$\delta = 0.22 R$. Two openings of width $W=0.222 R$ are attached 
opposite to each other at an
angle $\theta \in [0,\pi/2]$
relative to the line connecting the circle centers.
(b) Phase space, parameterized by the impact
parameter $s$ and the transverse momentum component $\sin\alpha$
of trajectories at the exterior circle.
The lower panels show the
phase space for open billiards with $\theta=0$ (c) and $\theta=\pi/2$
(d), for which we only record trajectories that are injected through the
openings
(until they leave the billiard again).
}
\label{fig1}
\end{figure}

A representative model for mixed regular and chaotic
classical dynamics is the
two-dimensional annular
billiard \cite{Bohigas}, Fig.\ \ref{fig1}(a), which consists of 
the region between two
circles with radii $R$, $r$, and eccentricity $\delta$.
Two leads (openings) of width $W$
are attached opposite to each other at an angle $\theta$ with respect to
the axis through the two circle centers.
The phase space can be parameterized by
the impact parameter $s$ and the transverse component
of the momentum $\sin\alpha$ (with $\alpha$
the angle of incidence) of trajectories that are reflected 
at the exterior circle. 
The phase space of the closed annular billiard 
for $r=0.6 R$, $\delta=0.22 R$ is shown in Fig.\ \ref{fig1}(b), which 
displays two regular 
whispering-gallery (WG) regions, a large regular island, 
neighboring satellite islands, and a chaotic sea.
Figs.\ \ref{fig1}(c) and (d) show the phase space of the open annular
billiard, which only includes trajectories of particles
that are injected into the billiard
through leads attached at $\theta=0$ and $\theta=\pi/2$, respectively
(the width of the leads is $W=0.222 R$).
The large island is well coupled to 
one of the openings for $\theta=0$,
and is completely decoupled from both openings at $\theta=0.5\pi$.
The chaotic sea and WG regions are well-coupled for arbitrary
position of the openings,
while the coupling of the satellite islands depends on $\theta$ and $W$.
In this way one can select regions in phase space
by varying $\theta$ and $W$.

We computed the dimensionless conductance $T={\rm tr}\, t^\dagger t$ and
the shot-noise suppression factor
${\cal F}=(2/N)\,{\rm tr}\, t^\dagger t(1-t^\dagger t)$
from the transmission 
matrix $t$ (the dimension of this matrix is given by
the number of channels $N$ in each lead)
\cite{Blanter_Phyrep}. The transmission matrix 
is obtained numerically
by the method 
of recursive Green functions \cite{Baranger_method}, for which space is 
discretized on a square lattice. In terms of the lattice constant
$a$, $R=144\,a$, $r=86.4\,a$, $\delta=31.7\,a$, and $W=32\,a$.
Energy $E$ will be measured in units $\hbar^2/(2 m a^2)$
and time in units of $2 m a^2/\hbar$, with $m$
the mass of the charge carriers. In these units
the mean level spacing $\Delta \simeq 0.0003$.
We will work in the energy window  $E\in(0.408,0.433)$, in which the
Fermi wavelength
$\lambda_{\rm F}\simeq 9.5\,a$, resulting in
$N= {\rm Int}(2W/\lambda_{\rm F})=6$.

\begin{figure}
\includegraphics[width=0.40\textwidth,height=0.20\textheight]{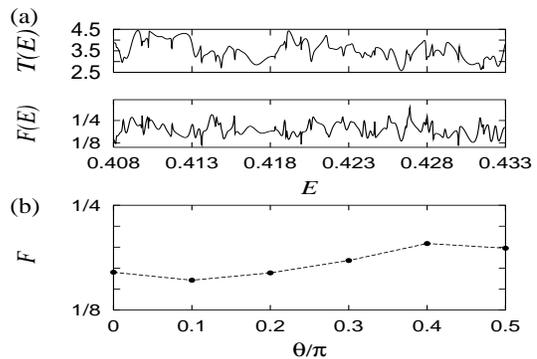}
\caption{
(a) Dimensionless conductance $T(E)$ 
and shot-noise suppression
factor ${\cal F}(E)$ for leads attached at $\theta=0.5\pi$.
(b) Energy-averaged suppression factor as a function of lead position $\theta$.
}
\label{fig2}
\end{figure}

The upper two panels of Fig.\ \ref{fig2} show $T$ and ${\cal F}$, as a
function of energy, 
for the case that the two leads are attached
at $\theta=\pi/2$.
The dependence of the 
energy-averaged suppression factor ${\cal F}$ on $\theta$
is shown in the lowest panel of  Fig.\ \ref{fig2}.
Our first observation is that
for all values of $\theta$, the suppression factor is smaller than the
universal value $1/4$ for fully chaotic motion.
Moreover, the
shot noise is more suppressed for the cases that one opening couples to 
the large regular island ($0<\theta\lesssim 0.3\pi$) than for the cases
in which the
regular islands is decoupled from the openings ($\theta \gtrsim 0.3\pi$).
This behavior indicates that electrons injected into the large regular island
contribute less to the shot noise.

In order to determine in detail which processes are responsible for the
additional shot-noise suppression, we introduce a function $f(t_{\rm
av})$ which
can be interpreted as the probability distribution function of
deterministic processes with dwell time $t_{\rm dwell}\simeq t_{\rm av}$.
This will be achieved by monitoring the rate of change of the shot noise,
\begin{equation}
f(t_{\rm av})=-(d/dt_{\rm av}){\cal F}(t_{\rm av})/{\cal F}_{\rm ch},
\label{eq:f}
\end{equation}
as we
successively replace the dynamics on time
scales $\gtrsim t_{\rm av}$ by fully chaotic dynamics (equivalent to
diffractive impurity scattering).
The function $f$ is normalized to the total shot-noise suppression
$\int_0^\infty dt_{\rm av} f(t_{\rm av})=1-{\cal F}/{\cal F}_{\rm ch}$,
taken relative to the universal value of chaotic dynamics.

The transmission matrix $t$ determining $T$ and ${\cal F}$
is a subblock of the scattering matrix
\begin{equation}
S=\left(\begin{array}{cc}r&t'\\t&r'\end{array}\right),
\end{equation}
where $r$ and $r'$ are reflection coefficients and the transmission
matrix $t'$ contains the same information as $t$.
The elimination of the system-specific details on time scales
$\gtrsim t_{\rm av}$
is achieved by averaging the scattering matrix $S$ over an energy range
$[E_0 - E_{\rm av}/2,E_0 + E_{\rm av}/2]$ of width  $E_{\rm av} =\hbar/t_{\rm
av}$ (which will be taken inside the total energy  range [0.408,0.433]
of our numerical simulation),
\begin{equation}
\overline{S}(E_{\rm av};E_0)= E_{\rm av}^{-1}
\int_{E_0 - E_{\rm av}/2}^{E_0 + E_{\rm av}/2}
 dE\,S(E).
\end{equation}
Here the information on processes with longer time scales
than $t_{\rm av}$ is lost, because it is encoded
into the short-range energy correlations (fluctuations) of the scattering matrix
\cite{vwz}, while the information on the dynamics on
shorter time scales than $t_{\rm av}$ modulates the scattering matrix on
larger energy scales and hence is retained. 
Chaotic processes
are then introduced to substitute the eliminated ones
by
coupling to an auxiliary chaotic system
with scattering matrix
$S_0$, taken from the appropriate circular ensemble
of random matrix theory (observing
the same symmetries as the original scattering matrix, as time-reversal
or spatial parities \cite{Baranger_Sym}), resulting in
\begin{eqnarray}
S'(E_{\rm av};E_0;S_0) = \overline{S}(E_{\rm av};E_0) + 
{\cal T}^\prime (1-S_0{\cal R})^{-1}S_0{\cal T}.
\label{smatrix}
\end{eqnarray}
The ensemble of scattering matrices (\ref{smatrix}) is the so-called
Poisson kernel \cite{Beenakker_RMP,Mello,Brouwer,Baranger_Poisson},
with $\overline S$ the so-called optical scattering matrix.
The coupling matrices ${\cal T}$, ${\cal T}^\prime$, and ${\cal R}$
must be chosen such that $S^\prime$ is a unitary matrix,
but the
invariance of the circular ensemble guarantees that
results do not depend on their specific choice.

\begin{figure}
\includegraphics[width=0.40\textwidth,height=0.24\textheight]{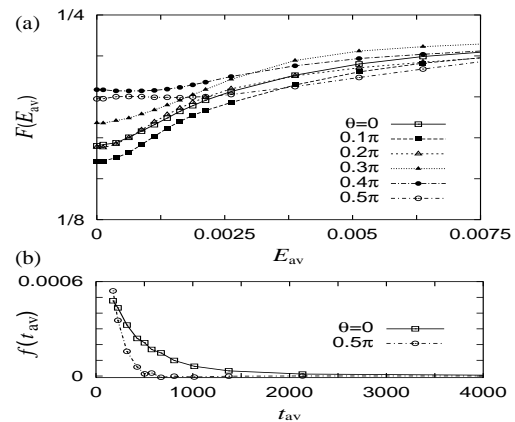}
\caption{
(a) Shot-noise suppression factor
${\cal F}$ as a function of the strength of diffractive
impurity scattering with a rate $\simeq E_{\rm av}/\hbar$,
for different positions $\theta$ of the leads. 
(b) Distribution function $f(t_{\rm av})$ of deterministic processes
for $\theta=0$ and $\theta=0.5\,\pi$. 
}
\label{fig3}
\end{figure}

We now can calculate the mean suppression
factor ${\cal F}(E_{\rm av})$ for fixed $E_{\rm
av}$ (and hence fixed $t_{\rm av}$), first
by averaging the noise within each Poisson kernel (fixing also $E_0$),
and then averaging these values over $E_0\in (0.408,0.433)$.
The result
for different positions of the
leads is shown in Fig.\ \ref{fig3}.
The value ${\cal F}(E_{\rm av}=0)$ is identical to the suppression factor
${\cal F}$ shown in Fig.\ \ref{fig2}(b), because the Poisson kernel
becomes a delta function at $S(E_0)$ as $\overline S$ approaches this unitary matrix
for $E_{\rm av}\to 0$.
For increasing $E_{\rm av}$ the suppression factor  ${\cal F}(E_{\rm av})$
approaches the universal value ${\cal F}_{\rm ch}$ of random-matrix
theory, because $\overline S=0$ in this limit.
The function ${\cal F}(E_{\rm av})$ is monotonically increasing
almost everywhere (negative values of the slope are within the
statistical uncertainty in the numerical simulation),
indicating 
that the noise is {\it enhanced} by the replacing the
system-specific deterministic details of the transport by the
indeterministic transport of random matrix theory.
Consequentially, as shown in Fig.\ \ref{fig3}(b),
the function $f(t_{\rm av})$ introduced in Eq.\
(\ref{eq:f}) is positive
and hence allows the interpretation of a probability
distribution function of deterministic scattering.

The most striking feature in Fig.\ \ref{fig3} is
that  ${\cal F}(E_{\rm av})$ rises quickly  in the three cases $\theta=0$,
$0.1\, \pi$,
$0.2\,\pi$, in which the leads couple to the large regular island in
phase space, while the
slope  of ${\cal F}(E_{\rm av})$ almost vanishes up to an energy $E_{\rm
indet} \approx 0.002$ for
$\theta=0.4\,\pi$, $0.5\,\pi$ (where the regular region is decoupled).
The distribution function $f(t_{\rm av})$ decays only slowly in
the former case, while vanishes almost identically beyond a time $t_{\rm
indet}\approx 500$ in the latter case.
This difference can be interpreted as follows.
Wave packets are trapped for long times
in regular regions, but show only little dispersion
because of the stability of regular trajectories \cite{cametti}.
Hence the scattering is deterministic and strongly affected
by introducing diffractive scattering even at these long time scales.
On the other hand, wave packets disperse quickly in the chaotic regions,
where the transport
is already
indeterministic for $t_{\rm av}\gtrsim t_{\rm indet}$, and this is
not modified by adding diffractive scattering.
The shot noise suppression for $\theta=0.4\,\pi$ and $\theta=0.5\pi$ 
hence arises from short chaotic trajectories with dwell times 
$\lesssim t_{\rm indet}$.
This is in accordance with the predictions of Ref.\ \cite{Agam};
chaotic quantum  transport is deterministic up to the Ehrenfest time
$(1/\lambda)\ln(2\pi R/\lambda_{\rm F}) \simeq 500$,
estimated from the Lyapunov exponent $\lambda\simeq
v_{\rm F}/R$ and the Fermi velocity $v_{\rm F}$.
Thus $f(t_{\rm av})$ reveals system-specific time scales as the
Ehrenfest time in chaotic regions and long dwell times in regular regions.
Moreover, $f(t_{\rm av})$ vanishes for
$t_{\rm av} \lesssim \min t_{\rm dwell} \simeq 50$
(not resolved in Fig.\ \ref{fig3}).

Our interpretation is confirmed by a semiclassical estimate which
discriminates the
contributions to the shot noise from different regions in phase
space.
We divide the phase space into Planck cells of area
$2W/(NR)$ and calculate the fluctuations in the
semiclassical occupation numbers
of these cells.
The average occupation number
$\overline{f}_j = T_{jL} f_L + T_{jR} f_R$
is obtained from the classical 
transmission probabilities  $T_{jL,R}$
from cell $j$ to the left (L)  and right (R) opening, respectively, where
$f_{L,R}$ is the Fermi distribution in the electronic reservoirs
attached to each opening.
Within the model of minimal correlations of Ref.\ \cite{Blanter_PRL},
the shot-noise suppression factor is then approximated as
\begin{equation}
{\cal F}_{\rm cl}  = 
\langle \overline{f}_j (1-\overline{f}_j) \rangle,
\label{clnoise}
\end{equation}
where $\langle\cdots\rangle$ denotes the average over the phase space cells
at the openings.

\begin{figure}
\includegraphics[width=0.40\textwidth,height=0.23\textheight]{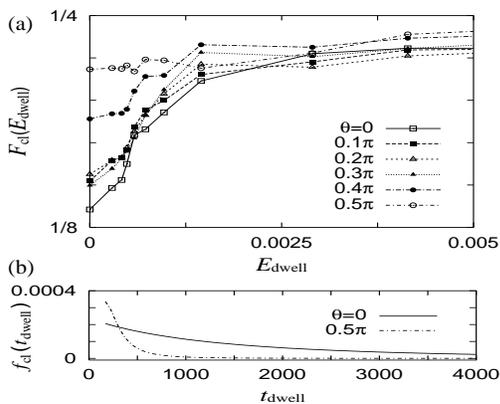}
\caption{
Semiclassical estimates of ${\cal F}$ and $f$
of Fig.\ \ref{fig3}, obtained by introducing indeterminism
into the classical dynamics beyond a given
dwell time $t_{\rm dwell}=\hbar/E_{\rm dwell}$. 
}
\label{fig4}
\end{figure}

Diffractive scattering can be introduced into the semiclassical theory
by replacing the
classical transmission probability of trajectories 
longer than a given  
dwell time $t_{\rm dwell}$
by the value $1/2$ of ergodic dynamics.
[The universal value of shot noise for chaotic dynamics follows from
Eq.\ (\ref{clnoise}) by inserting
the ergodic value $\overline{f}_j=1/2$.]
Figure \ref{fig4} shows the resulting
${\cal F}_{\rm cl}(E_{\rm dwell})$ and $f_{\rm cl}(t_{\rm dwell})$,
with $E_{\rm dwell}=\hbar/t_{\rm dwell}$.
The same relative time scales 
for the different positions of the leads are found as in the exact results
in Fig.\ \ref{fig3}.
The agreement is reasonable
given that we are not deep in the semiclassical limit
and accounting for the fact that the model of minimal correlations in Ref.\
\cite{Blanter_PRL} originally was devised for chaotic dynamics.
The suppression factor can be further
decomposed into contributions of different regions in phase
space.
For $\theta=0$, ${\cal F}_{\rm cl}(E_{\rm dwell})$ initially  increases
due to the large regular island and then due to deterministic 
processes in the chaotic sea, while for $\theta=0.5\pi$ it is
constant up to $E_{\rm dwell} \simeq E_{\rm indet}$.
Unlike the large regular island, the regular WG regions
(which we have ignored so far in our discussion) contribute to shot 
noise with the universal value, which is explained
by the fact that for the present system
parameters Planck cells do not yet
resolve reflected from transmitted trajectories which are injected
into these regions.
One can see from the model of minimal correlations, Eq.\ (\ref{clnoise}),
that shot noise eventually will be suppressed also in the WG regions
when the semiclassical limit is approached.
The same holds for the
hierarchical structure of small regular regions embedded in the
chaotic part of phase space. A pronounced shot noise suppression should
arise once that these trapping regions are resolved, because in the vicinity
of stable structures the Ehrenfest time depends only algebraically
on $\hbar$ \cite{cametti}.

In conclusion, we have demonstrated that shot noise is a 
measure of the amount of deterministic transport through
generic quantum dots.
The dynamical analysis developed in this work may be used to
study time scales in a large variety of systems.
Experimentally, these time scales could be probed by
measuring shot noise while tuning the indeterministic scattering rate
by adjusting a gate voltage \cite{toyoda}.

We thank S. Tomsovic for useful discussions 
and P. W. Brouwer for advice on implementing the Poisson kernel.

\bibliographystyle{apsrev}


\end{document}